\documentclass[aps,twocolumn,prc,showpacs]{revtex4}
\usepackage{graphicx}
 \def\be{\begin{equation}}
 \def\ee{\end{equation}}
 \def\bea{\begin{eqnarray}}
 \def\eea{\end{eqnarray}}
 
 \def\l{\left}
 \def\r{\right}
 \def\Im{{\rm Im}}
 \def\Re{{\rm Re}}
 \def\bm#1{\mbox{\boldmath$#1$}}
 \def\gsim{\mathrel{\rlap{\lower0.2em\hbox{$\sim$}}\raise0.2em\hbox{$>$}}}
 \def\ksim{\mathrel{\rlap{\lower0.2em\hbox{$\sim$}}\raise0.2em\hbox{$<$}}}

\begin{document}
\title{The hot non-perturbative gluon plasma is an almost ideal colored liquid}

\author{A.~Peshier and W.~Cassing}

\affiliation{
 Institut f\"{u}r Theoretische Physik,
 Universit\"{a}t Giessen, 35392 Giessen, Germany}
\date{\today}

\begin{abstract}
We study properties of a gluon plasma above the critical temperature $T_c$
in a generalized quasi-particle approach with a Lorentz spectral function.
The model parameters are determined by a fit of the entropy $s$ to lattice
QCD data.
The effective degrees of freedom are found to be rather heavy and of a
sizeable width.
With the spectral width being closely related to the interaction rate, we
find a large effective cross section, which is comparable to the typical
distance squared of the quasiparticles.
This suggests that the system should be viewed as a liquid as also
indicated by an estimate of the plasma parameter $\Gamma$.
Furthermore, within the quasiparticle approach we find a very low viscosity
to entropy ratio, $\eta/s \sim 0.2$ for $T > 1.05 T_c$, supporting the recent
conjecture of an almost ideal quark-gluon liquid seen at RHIC.
\end{abstract}

\maketitle


The formation of a quark-gluon plasma (QGP) and its transition to interacting
hadronic matter -- as occurred in the early universe -- has motivated a large
community for more than two decades (cf.\ \cite{QM01} and Refs.\ therein).
However, the complexity of the dynamics in ultrarelativistic nucleus-nucleus
collisions -- producing high density matter for short time scales -- has not
been fully unravelled and  many properties of the new phase are still under
debate \cite{SQM04}.  In central Au+Au collisions energy densities are reached
at the Relativistic Heavy Ion collider (RHIC) that are far above the critical
energy density $e_c \sim 1\,$GeV/fm$^3$ for a phase transition to a QGP as
expected from lattice QCD calculations \cite{Karsch}.
A strong radial expansion and elliptic flow of hadrons, furthermore, point
towards an early generation of pressure and a high interaction rate in the
`new phase' \cite{Brat03}.

The latter observables are severely underestimated in conventional
string/hadron transport models \cite{Cassing03,Brat04,Cassing04}, however
hydrodynamical approaches do quite well in describing (at midrapidity) the
collective properties of the system for low and moderate transverse momenta 
\cite{Heinz}. The picture thus emerges that the medium created in 
ultrarelativistic nucleus-nucleus collisions for a couple of fm/c interacts
more strongly than hadron/string matter, and it exhibits collective properties
that resemble those of a liquid of low shear viscosity $\eta$ \cite{Shuryak}.
In fact, viscous hydrodynamical calculations indicate a very low viscosity to
entropy ratio, $\eta/s \approx 0.1...0.2$ \cite{Teaney}.
This picture is substantially different from the expectation of a weakly
coupled colored plasma. There is a variety of models that address the
properties of this `new matter'. It might be some kind of i) `epoxy'
\cite{GerryEd}, i.e.\ a system of resonant or bound gluonic states with large
scattering length, ii) a system of chirally restored mesons, instanton
molecules or equivalently giant collective modes \cite{GerryRho}, iii) a
system of colored bound states of quarks $q$ and gluons $g$, i.e.\ $gq$, $qq$,
$gg$ etc.\ \cite{Eddi}.

In this letter we will provide quantitative arguments that strongly 
interacting matter in a certain temperature range above $T_c$ is in a 
liquid phase. Our arguments are based on a generalized quasiparticle 
description of the system taking into account the spectral width $\gamma$
\cite{Andre04} in addition to the quasiparticle mass
\cite{BlaizIR,Peshi,pQP,RebhaR,Toneev}.
The model parameters are adjusted to non-perturbative results of lattice
calculations. Since the width is closely related to the interaction rate
we can then estimate relevant transport properties, such as the
effective cross section and the shear viscosity, for temperatures
near $T_c$.
Since lattice `data' are more precise for quenched QCD we focus here on
pure gluonic systems. We expect, however, similar results for full
QCD as argued below.
The quantities we address are dominated by `hard' momenta of the order of the
temperature $T$. Accordingly, the quasiparticle properties we are interested
in are related to the gluon propagator at hard momentum scales.
In this line we take into account $d_g = 2 (N^2_c-1) = 16$ transverse gluons,
and neglect Landau-damping contributions as well as the collective
longitudinal modes whose spectral strength is suppressed for larger momenta
\cite{BlaizIR, Peshi}.

In order to adjust the quasiparticle properties we first consider 
thermodynamic bulk properties within the $\Phi$-derivable formalism
\cite{LuttiW}, which yields consistent resummed approximations \cite{Baym}.
To leading-loop order, which is expedient for large coupling as argued in
\cite{Andre04}, the entropy follows directly from the quasiparticle
propagator, cf.\ Ref.~\cite{BlaizIR},
\be
  s^{dqp}
  =
  - d_g\!\int\!\!\frac{d \omega}{2 \pi} \frac{d^3p}{(2 \pi)^3}
  \frac{\partial n}{\partial T}
   \l( \Im\ln(-\Delta^{-1}) + \Im\Pi\,\Re\Delta \r)\!,
  \label{eq: s dqp}
\ee
where $n(\omega) = (\exp(\omega/T)-1)^{-1}$ denotes the Bose distribution
function.
We note that in the context of Fermi liquid theory a corresponding
approximation is called the {\em dynamical quasiparticle} (dqp) entropy
\cite{sFctnl}.
In principle, the resummed propagator $\Delta = (\omega^2-\bm p^2 - \Pi)^{-1}$
is to be calculated from a 1-loop Schwinger-Dyson equation.
To proceed at this point, however, we use a physically motivated {\em Ansatz},
assuming a Lorentzian spectral function,
\be
 \rho(\omega)
 =
 \frac\gamma{ E} \l(
   \frac1{(\omega-E)^2+\gamma^2} - \frac1{(\omega+E)^2+\gamma^2}
 \r) .
 \label{eq: rho}
\ee
With the convention $E^2(\bm p) = \bm p^2+M^2-\gamma^2$, the parameters $M^2$
and $\gamma$ are directly related to the real and imaginary parts of the
corresponding (retarded) self-energy, $\Pi = M^2-2i\gamma\omega$.
At this point we emphasize that the entropy functional (\ref{eq: s dqp}) is
not restricted to `strict' quasiparticles, i.e., $\gamma$ need not to be small
compared to the typical energy.
Following the models \cite{pQP} we parameterize the quasiparticle mass in the
gauge invariant and momentum independent asymptotic form
\be
 M^2 = \frac{N_c}6\, g^2 T^2 \, ,
 \label{eq: M2}
\ee
with $N_c = 3$ and the running coupling
\be
 g^2(T) = \frac{48\pi^2}{11N_c \ln(\lambda(T-T_s)/T_c)^2} \, ,
 \label{eq: g2}
\ee
which permits an enhancement near $T_c$ \cite{pQP,Rafelski}.
Likewise, we parameterize the width in the form $\gamma \sim g^2 T \ln g^{-1}$
\cite{Pisar89LebedS} or, equivalently, in terms of $M$ \cite{Andre04},
\be
  \gamma
  =
  \frac3{4\pi}\, \frac{M^2}{T^2} \, T \ln\frac{c}{(M/T)^2} \, ,
 \label{eq: gamma}
\ee
where $c$ is related to the magnetic sector of QCD.
We note that $M$ and $\gamma$, as parameterizations of the complex-valued
self-energy $\Pi$ at the relevant large momenta near the `mass shell', are 
not related by a dispersion relation.

In the upper part of Fig.~1 we compare the lattice results \cite{CCPACS} to the
quasiparticle entropy; the fitted parameters are $\lambda = 2.42$, $T_s= 0.46$
and $c=14.4$. We also display the interaction measure $e-3p$ (the pressure $p$
and the energy density $e$ are evaluated by thermodynamic relations,
cf.~\cite{pQP}), which is particularly sensitive to interaction
effects. We emphasize that the remarkable agreement with the lattice data is
non-trivial because the functional relation between $\gamma$ and $M$ is fixed,
cf.~(\ref{eq: gamma}).
\begin{figure}[ht]
  \centerline{\includegraphics[scale=0.72]{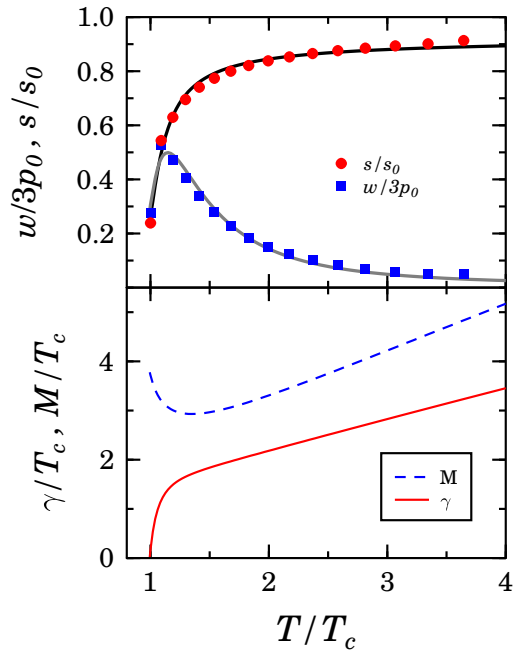}}
  \caption{The entropy $s$ and the interaction measure $w=e-3p$, in units
  of the Stefan-Boltzmann limits $s_0$ and $p_0$, from our quasiparticle model
  in comparison to lattice calculations \cite{CCPACS}.
  The lower part shows the adjusted mass $M$ and width $\gamma$.}
\end{figure}

The adjusted quasiparticle mass and width are displayed in the lower part of
Fig.~1. The quasiparticles are rather heavy in line with direct lattice
calculations \cite{Petre}.
For $T \gsim 1.05 T_c$ the width is sizeable, reaching more than 50\% of
the mass in a large temperature range.
The picture of the strongly interacting plasma is, thus, a system of massive
excitations with a large collisional width or short mean-free path -- opposite
to the original concept of {\em narrow} quasiparticles.
Near $T_c$, however, the width is close to zero.
Although we have parameterized $\gamma$ in the `perturbative' form
(\ref{eq: gamma}) we expect that the inferred temperature dependence is
generic: near $T_c$ the width has to be small due to the small entropy at
$T \approx T_c$ \cite{Andre04}. This is in line with a critical slowing
down near a phase transition.
Away from $T_c$ one has to expect a large width due to the strong coupling
and increasing reaction rates.
The physical processes contributing to the width are then
$gg \leftrightarrow gg$ scatterings as well as splitting and fusion reactions
$gg \leftrightarrow g$ or $gg \leftrightarrow ggg$ etc.
Summing up the elastic and inelastic channel (and neglecting Bose-enhancement
for the final states) we end up with the total binary reaction rate
\bea \!\!\!
 \frac{d N_{coll}}{dV dt}
 &=&
 {\rm \tilde Tr}_{P_1} {\rm \tilde Tr}_{P_2} \,
 2 \sqrt{\lambda(s,P_1^2,P_2^2)}\; \sigma_{tot}(P_1,P_2)
 \nonumber \\
 &=&
 \langle \sigma \rangle\,
 {\rm \tilde Tr}_{P_1} {\rm \tilde Tr}_{P_2} \,
 2 \sqrt{\lambda(s,P_1^2,P_2^2)}
 =:
 \langle \sigma \rangle \, I_2,
 \label{eq: rate}
\eea
where $\lambda(x,y,z) = (x-y-z)^2 - 4yz$ and $s = (P_1+P_2)^2$.
In (\ref{eq: rate}) we have introduced the shorthand notation
\be
 {\rm \tilde Tr}_P \cdots
 =
 d_g\!\int\!\!\frac{d \omega}{2 \pi} \frac{d^3p}{(2 \pi)^3}\,
 2\omega\, \rho(\omega)\, \Theta(\omega) \Theta(P^2)\, n(\omega) \cdots \,
 \label{eq: short}
\ee
for the thermally weighted trace over the quasiparticle degrees of freedom.
The $\Theta(P^2)$ function ensures that only time-like reaction processes
are taken into account in Eq.~(\ref{eq: rate}).
The interaction rate, on the other hand, is also related to the imaginary part 
of the self-energy; with a similar factorization as in Eq.~(\ref{eq: rate}),
$d N_{coll}/ dV dt = \gamma N_+$.
Here the particle density $N_+ = {\rm {\tilde Tr}}\, 1$ is the time-like
part of the integrated distribution function.
The resulting effective total cross section,
\be
  \langle\sigma \rangle = \gamma N_+/I_2 \, ,
  \label{eq: sigma}
\ee
is displayed in the top part of Fig.~2 as a function of $T$ (using
$T_c = 0.26\,$GeV for quenched QCD). It rises from $\approx 0$ at $T = T_c$ to
about $20\,$mb at $T \approx 1.1T_c$, and drops again at higher temperatures.
We note that similarly large values for parton cross sections have been used in
the phenomenological studies in Ref.~\cite{Molnar}. These cross sections are
larger by an order of magnitude than typical perturbative estimates for $gg$
scattering (cf.\ the hatched band in the top part of Fig.~2) where the Debye
mass $m_{\rm Debye}$ is used as an infrared cutoff. However, for strongly
coupled plasmas the Debye mass may not be the proper regulator as pointed
out by Thoma \cite{Thoma}.
\begin{figure}[ht]
   \centerline{\includegraphics[scale=0.72]{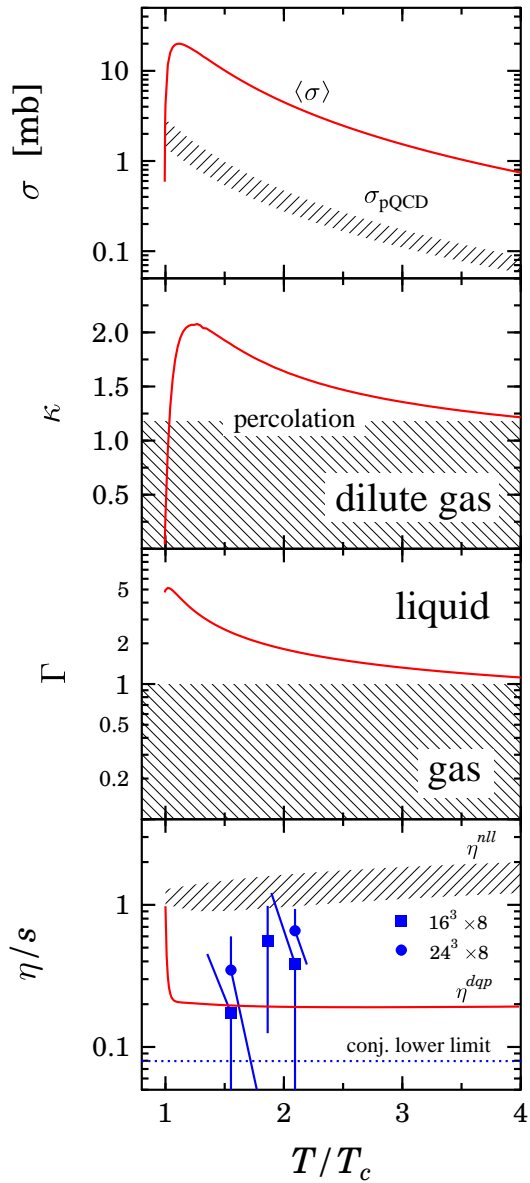}}
   \caption{Upper part: The effective total cross section
   $\langle\sigma\rangle$, Eq.~(\ref{eq: sigma}), in comparison to the
   perturbative estimate $\sigma_{\rm pQCD} = 9\pi\alpha^2/2m_{\rm Debye}^2$.
   2nd part from top: The percolation measure $\kappa$, Eq.~(\ref{eq: kappa});
   the critical value is $\kappa_c =1.18$ \cite{Satz}.
   3rd part from top: The plasma parameter $\Gamma$, Eq.~(\ref{eq: Gamma}).
   Bottom part: The ratio of shear viscosity to entropy in our quasiparticle
   model in comparison to the lattice calculation \cite{lattice2} as well as
   to the next-to-leading log (nll) result \cite{Arnold}.
   Note that $\eta/s \approx 0.1 \ldots 0.2$ was estimated from hydrodynamical
   fits to RHIC data \cite{Teaney}.}
\end{figure}

A question of particular interest is the phase structure of the strongly
interacting system, which in Refs.~\cite{Thoma,Shuryak} was surmised a liquid.
In our quasiparticle approach we can address this question quantitatively by
comparing the effective cross section to the typical distance squared of the
quasiparticles to obtain information about critical clustering (percolation)
\cite{Satz}. If
\be
  \kappa = \langle\sigma\rangle {N_+^{2/3}}
  \label{eq: kappa}
\ee
is lower than the critical percolation parameter $\kappa_c \approx 1.18$
\cite{Satz} the system is in the kinetic (dilute gas) regime while for
$\kappa > \kappa_c$ percolation sets in and multi-particle interactions take
over as characteristic for a liquid or solid with attractive interactions.
As shown in the second part of Fig.~2, $\kappa$ is larger than $\kappa_c$ for
the temperature range $1.05 \ldots 4T_c$.
This suggests that the QCD plasma, up to rather large energy densities, is
in a liquid/solid phase.

In order to distinguish a liquid from, possibly, a solid phase we consider
the plasma parameter \cite{Thoma}.
Assuming equal (color) magnetic and electric energies we estimate
\be
  \Gamma
  =
  2\frac{N_c g^2}{4\pi\, N^{-1/3}}\, \frac1{\langle T_{kin}\rangle} \, ,
  \label{eq: Gamma}
\ee
with $N$ being the full particle density (including the space-like
contributions), and the average kinetic energy
\be
  \langle T_{kin} \rangle
  =
  N_+^{-1}\, {\rm {\tilde Tr}} \l( \omega - \sqrt{p^2} \r) .
\ee
Empirically it is well established for various attractive interactions that 
systems with $1 \ksim \Gamma \ksim 100$ are in a liquid phase \cite{Liquid}.
Since the `critical' values may slightly vary for different systems we
emphasize for the present case that the criteria $\Gamma_c = 1$ for a
gas-liquid phase transition and $\kappa = \kappa_c$ for percolation are
met at the same temperature, cf. Fig.~2.
Thus our results suggest that the gluon plasma is a liquid for
temperatures between $T_c$ and $4T_c$ .

An important property of this gluon liquid is its shear viscosity $\eta$.
For weak coupling, it has been calculated in a transport approach from a
Boltzmann equation to next-to-leading log order \cite{Arnold},
$\eta^{nll} \sim T^3/(g^4 \ln g^{-1})$.
However, for strongly coupled systems such an approach (assuming {\em
narrow} quasiparticles) might be questionable and the Kubo formalism
more appropriate instead \cite{Edward2}. Here the viscosity is
evaluated from the slope of the Fourier transform of the spectral
function $\langle [T_{ij}(x), T_{ij}(y)] \rangle$ for $\omega
\rightarrow 0$, where $T_{ij}$ denotes the traceless part of the spatial
stress tensor. At 1-loop order (and neglecting the longitudinal
contributions) this corresponds directly to our quasiparticle
picture which yields, cf.~\cite{Aarts},
\be
  \eta^{dqp}
  =
  -\frac{d_g}{60}\!\int\!\!\frac{d\omega}{2\pi} \frac{d^3p}{(2 \pi)^3}\,
   \frac{\partial n}{\partial \omega}\, \rho^2(\omega)
   \l[ 7\omega^4 - 10\omega^2 \bm p^2 + 7\bm p^4 \right] .
  \label{eq: eta}
\ee
Note that in the perturbative limit Eq.~(\ref{eq: eta})
($\sim T^4/\gamma \sim T^3/(g^2 \ln g^{-1})$) does not approach
$\eta^{nll} \sim T^3/(g^4 \ln g^{-1})$.
This comes about as follows: The shear viscosity $\eta$ is inversely
proportional to the transport cross section $\sigma_{trans}$ in
which the total scattering rate is weighted by $(1-\cos\theta)$ with
$\theta$ denoting the scattering angle \cite{Thoma2}.
In the weak coupling limit the gluon scattering is strongly forward
peaked with a low gain in transverse momentum.
This implies that many scatterings, corresponding to ladder diagrams in the
Kubo formalism, have to be resummed in order to achieve a significant
transverse momentum deflection and transport cross section.
However, in the case of strong coupling the transport mean-free-path
$\lambda_{trans} \sim 1/(\sigma_{trans} N_+)$ is expected to become 
comparable to the total mean-free-path $\lambda \sim 1/(\sigma N_+)$, as 
taken into account by Eq.~(\ref{eq: eta}).
Thus, while a resummation of ladder diagrams is beyond the scope of the
quasiparticle model, it should yield a useful approximation near $T_c$
\cite{new}.

In the bottom part of Fig.~2 we display the ratio of the shear viscosity
to entropy. Although the lattice results \cite{lattice2} still have large
uncertainties, they are distinctly smaller than the (extrapolation of the)
next-to-leading log result \cite{Arnold}.
For $T \gsim 1.05 T_c$ our quasiparticle result is almost constant,
$\eta^{dqp}/s \approx 0.2$  -- in good agreement with the estimate
$\eta/s \approx 0.1...0.2$ from hydrodynamical fits to RHIC data \cite{Teaney}.
The fast increase near $T_c$, which is related to the characteristic
temperature dependence of $\gamma$ and thus also of $\langle \sigma \rangle$,
can be seen as a precursor of the phase transition.
We mention that the conjectured lower limit \cite{Starinets},
$\eta/s \ge 1/(4\pi)$, is approximately obtained for $T \approx 1.1T_c$ when
calculating $\eta$ in a non-relativistic limit of massive quasiparticles and
assuming isotropic transport cross sections \cite{new}. Since $\sigma_{trans}
\leq \sigma$ this estimate should give a lower bound for the viscosity.
In conclusion, the quasiparticle model gives the picture of an almost
ideal gluon liquid in the relevant temperature range.

The extension of the quasiparticle approach to the physical case (full QCD) 
is straightforward; besides a change of $T_c$, quarks and gluons have the 
same quasiparticle properties up to group factors \cite{Peshi04b}.
The consequences for ultra-relativistic heavy-ion collisions at RHIC become
immediately clear: The large cross sections imply a rapid thermalization of
the initial configuration once the initial hard scatterings have produced
a high density of mini-jets. The latter `pre-equilibrium' processes
happen on a scale of $t_{pre} = 2R_A/\gamma$ $\leq 0.14$ fm/c for
top RHIC energies such that the equilibration time is essentially
governed by the $gg \leftrightarrow g$, $gg \leftrightarrow gg$,
$gg \leftrightarrow ggg$ and $gg \leftrightarrow gggg$ processes as
also suggested in \cite{Carsten}.
After approximate thermalization -- of order 0.5 to 1 fm/c -- the
system behaves like an almost ideal massive colored parton liquid
and exhibits a large pressure. This early pressure is responsible for
the transverse flow of hadrons, and the large cross sections result in
an almost complete suppression of far-side jets in central collisions.
\\[3mm]
{\bf Acknowledgments:}
The authors would like to thank C.~Greiner, S.~Leupold, R.~Pisarski,
D.~Teaney, M.~Thoma, and X.N.~Wang for helpful discussions and valuable
suggestions.
This work was supported by BMBF.


\begin{thebibliography}{99}
\bibitem{QM01}
    {\it Quark Matter 2002}, Nucl.\ Phys.\ A715 (2003) 1;
    {\it Quark Matter 2004}, J.\ Phys.\ G 30 (2004) S633.

\bibitem{SQM04} {\it Strange Quark Matter 2003}, J.\ Phys.\ G 30 (2004) 1.

\bibitem{Karsch} F.~Karsch {\it et al.}, Nucl.\ Phys.\ B605 (2001) 579.

\bibitem{Brat03} H.~Weber {\it et al.}, Phys.\ Rev.\ C 67 (2003) 014904.

\bibitem{Brat04} E.L.~Bratkovskaya {\it et al.},
 Phys.\ Rev.\ C 67 (2003) 054905; Phys.\ Rev.\ C 69 (2004) 054907.

\bibitem{Cassing03} W.~Cassing, K.~Gallmeister, C.~Greiner,
 Nucl.\ Phys.\ A735 (2004) 277.

\bibitem{Cassing04} K.~Gallmeister, W.~Cassing,
 Nucl.\ Phys.\ A748 (2005) 241.

\bibitem{Heinz} P.~Kolb, U.~Heinz, nucl-th/0305084.

\bibitem{Shuryak} E.V.~Shuryak, Prog.\ Part.\ Nucl.\ Phys.\ 53 (2004) 273.

\bibitem{Teaney} D.A.~Teaney, J.\ Phys.\ G 30 (2004) S1247.

\bibitem{GerryEd} G.E.~Brown, C.-H.~Lee, M.~Rho, E.V.~Shuryak,
 Nucl.\ Phys.\ A740 (2004) 171.

\bibitem{GerryRho} G.E.~Brown, C.-H. Lee, M.~Rho,
 Nucl.\ Phys.\ A747 (2005) 530.

\bibitem{Eddi} E.V.~Shuryak, I.~Zahed, Phys.\ Rev.\ D 70 (2004) 054507.

\bibitem{Andre04} A.~Peshier, Phys.\ Rev.\ D 70 (2004) 034016.

\bibitem{BlaizIR} J.\,P.~Blaizot, E.~Iancu, A.~Rebhan,
 Phys.\ Rev.\ D 63 (2001) 065003.

\bibitem{Peshi} A.~Peshier, Phys.\ Rev.\ D 63 (2001) 105004.

\bibitem{RebhaR} A.~Rebhan, P.~Romatschke, Phys.\ Rev.\ D 68 (2003) 025022.

\bibitem{pQP}
 A.~Peshier, B.~K{\"a}mpfer, O.\,P.~Pavlenko, G.~Soff,
 Phys.\ Rev.\ D 54 (1996) 2399;
 P.~Levai, U.~Heinz, Phys.\ Rev.\ C 57 (1998) 1879;
 A.~Peshier, B.~K{\"a}mpfer, G.~Soff, Phys.\ Rev.\ C 61 (2000) 045203,
 Phys.\ Rev.\ D 66 (2002) 094003.

\bibitem{Toneev} Yu.B.~Ivanov, V.V.~Skolov, V.D.~Toneev,
 Phys.\ Rev.\ D 71 (2005) 014005.

\bibitem{LuttiW} J.\,M.\ Luttinger, J.\,C.\ Ward, Phys.\ Rev.\ 118 (1960) 1417.

\bibitem{Baym} G.~Baym, Phys.\ Rev.\ 127 (1962) 1391.

\bibitem{sFctnl} G.\,M.~Carneiro, C.\,J.~Pethick, Phys.\ Rev.\ B 11 (1975) 1106.

\bibitem{Rafelski} J.~Letessier, J.~Rafelski,
  Phys.\ Rev.\ C 67 (2003) 031902.

\bibitem{Pisar89LebedS}
 R.\,D.~Pisarski, Phys.\ Rev.\ Lett.\ 63 (1989) 1129;
 V.\,V.~Lebedev, A.\,V.~Smilga, Ann.\ Phys.\ (N.Y.) 202 (1990) 229.

\bibitem{CCPACS}
 M.~Okamoto {\em et al.}, Phys.\ Rev.\ D 60  (1999) 094510.

\bibitem{Petre}
  P.~Petreczky {\em et al.}, Nucl.\ Phys.\ (Proc.\ Suppl.) B106 (2002) 513.

\bibitem{Molnar} D.~Molnar, M.~Gyulassy, Nucl.\ Phys.\ A698 (2002) 379.

\bibitem{Thoma} M.H.~Thoma, J.\ Phys.\ G 31 (2005) L7; 539.

\bibitem{Satz} H.~Satz, Nucl.\ Phys.\ A661 (1999) 104c.

\bibitem{Liquid} S.~Ichimaru, Rev.\ Mod.\ Phys.\ 54 (1982) 1017.

\bibitem{Arnold} P.~Arnold, G.D.~Moore, L.G.~Yaffe,
 JHEP 0305 (2003) 051; 0301 (2003) 030.

\bibitem{Edward2} B.A.~Gelman, E.V.~Shuryak, I.~Zahed, nucl-th/0410067.

\bibitem{Aarts} G.~Aarts, J.M.~Martinez Resco,
 JHEP 0204 (2002) 053; JHEP 0402 (2004) 061.

\bibitem{Thoma2} M.H.~Thoma, Phys.\ Rev.\ D49 (1994) 451.

\bibitem{new} A.~Peshier, W.~Cassing (to be published).

\bibitem{lattice2} A.~Nakamura, S.~Sakai, 
 Phys.\ Rev.\ Lett.\ 94 (2005) 072305.

\bibitem{Starinets} P.~Kovtun, D.T.~Son, A.O.~Starinets, 
 Phys.\ Rev.\ Lett.\ 94 (2005) 111601.

\bibitem{Peshi04b} A.~Peshier, 
 J.\ Phys.\ G 31 (2005) S371.

\bibitem{Carsten} Z. Xu, C. Greiner, hep-ph/0406278.

\end{thebibliography}
\end{document}